\documentclass[letter,prd,aps,twocolumn,floatfix,superscriptaddress]{revtex4}

\usepackage{amssymb,amsmath,graphicx}
\usepackage{natbib}
\usepackage{hyperref}
\makeatletter
\renewcommand*{\p@section}{\S\,}
\renewcommand*{\p@subsection}{\S\,}

\makeatother
\usepackage[usenames]{color}

\def\jnl@style{\it}
\def\aaref@jnl#1{{\jnl@style#1}}

\def\aaref@jnl#1{{\jnl@style#1}}

\def\aj{\aaref@jnl{AJ}}                   
\def\araa{\aaref@jnl{ARA\&A}}             
\def\apj{\aaref@jnl{ApJ}}                 
\def\apjl{\aaref@jnl{ApJ}}                
\def\apjs{\aaref@jnl{ApJS}}               
\def\ao{\aaref@jnl{Appl.~Opt.}}           
\def\apss{\aaref@jnl{Ap\&SS}}             
\def\aap{\aaref@jnl{A\&A}}                
\def\aapr{\aaref@jnl{A\&A~Rev.}}          
\def\aaps{\aaref@jnl{A\&AS}}              
\def\azh{\aaref@jnl{AZh}}                 
\def\baas{\aaref@jnl{BAAS}}               
\def\jrasc{\aaref@jnl{JRASC}}             
\def\memras{\aaref@jnl{MmRAS}}            
\def\mnras{\aaref@jnl{MNRAS}}             
\def\pra{\aaref@jnl{Phys.~Rev.~A}}        
\def\prb{\aaref@jnl{Phys.~Rev.~B}}        
\def\prc{\aaref@jnl{Phys.~Rev.~C}}        
\def\prd{\aaref@jnl{Phys.~Rev.~D}}        
\def\pre{\aaref@jnl{Phys.~Rev.~E}}        
\def\prl{\aaref@jnl{Phys.~Rev.~Lett.}}    
\def\pasp{\aaref@jnl{PASP}}               
\def\pasj{\aaref@jnl{PASJ}}               
\def\qjras{\aaref@jnl{QJRAS}}             
\def\skytel{\aaref@jnl{S\&T}}             
\def\solphys{\aaref@jnl{Sol.~Phys.}}      
\def\sovast{\aaref@jnl{Soviet~Ast.}}      
\def\ssr{\aaref@jnl{Space~Sci.~Rev.}}     
\def\zap{\aaref@jnl{ZAp}}                 
\def\nat{\aaref@jnl{Nature}}              
\def\iaucirc{\aaref@jnl{IAU~Circ.}}       
\def\aplett{\aaref@jnl{Astrophys.~Lett.}} 
\def\apspr{\aaref@jnl{Astrophys.~Space~Phys.~Res.}}
\def\bain{\aaref@jnl{Bull.~Astron.~Inst.~Netherlands}} 
\def\fcp{\aaref@jnl{Fund.~Cosmic~Phys.}}  
\def\gca{\aaref@jnl{Geochim.~Cosmochim.~Acta}}   
\def\grl{\aaref@jnl{Geophys.~Res.~Lett.}} 
\def\jcp{\aaref@jnl{J.~Chem.~Phys.}}      
\def\jgr{\aaref@jnl{J.~Geophys.~Res.}}    
\def\jqsrt{\aaref@jnl{J.~Quant.~Spec.~Radiat.~Transf.}}
\def\memsai{\aaref@jnl{Mem.~Soc.~Astron.~Italiana}}
\def\nphysa{\aaref@jnl{Nucl.~Phys.~A}}   
\def\physrep{\aaref@jnl{Phys.~Rep.}}   
\def\physscr{\aaref@jnl{Phys.~Scr}}   
\def\planss{\aaref@jnl{Planet.~Space~Sci.}}   
\def\procspie{\aaref@jnl{Proc.~SPIE}}   

\newcommand{\Alfven}{Alfv\`en}

\begin{document}

\date{\today}
\title{Magnetized Neutron Stars With Realistic Equations
      of State and Neutrino Cooling}

%
%
\author{David Neilsen}
\affiliation{Department of Physics and Astronomy, Brigham Young University, Provo, Utah 84602, USA}
\author{Steven L. Liebling}
\affiliation{Department of Physics, Long Island University, Brookville, New York 11548, USA}
\author{Matthew Anderson}
\affiliation{Pervasive Technology Institute, Indiana University, Bloomington, IN 47405, USA}
\author{Luis Lehner}
\affiliation{Perimeter Institute for Theoretical Physics,Waterloo, Ontario N2L 2Y5, Canada}
\author{Evan O'Connor}
\affiliation{Canadian Institute for Theoretical Astrophysics, Toronto, Ontario M5S 3H8, Canada}
\author{Carlos Palenzuela}
\affiliation{Canadian Institute for Theoretical Astrophysics, Toronto, Ontario M5S 3H8, Canada}

\begin{abstract}
We incorporate realistic, tabulated  equations of state into fully relativistic
simulations of magnetized neutron stars along with a neutrino leakage scheme which
accounts for cooling via neutrino emission. Both these improvements
utilize open-source code (GR1D) and tables from \href{http://stellarcollapse.org}{stellarcollapse.org}.
Our implementation makes use of a novel method for the calculation of
the optical depth which simplifies its use with distributed adaptive mesh
refinement. 
We present various tests 
with and without magnetization
and preliminary results both from single stars and from
the merger of a binary system.
\end{abstract}

\maketitle


\section{Introduction}\label{introduction}
Non-vacuum, compact, binary systems (i.e. two neutron stars or 
a black hole paired with a neutron star)
provide one of the most exciting laboratories to test fundamental aspects
of diverse physics. For instance, gravitational wave observations,
such as those expected from the detectors LIGO, VIRGO, and KAGRA~\citep{Abbott:2007kv,2011CQGra..28k4002A,Somiya:2011np},
can: shed light on the demographics of such binaries,  reveal aspects of
the equation of state~(EoS) at 
nuclear densities~\cite{Read:2009yp,Lackey:2011vz,Read:2013zra}, serve as stringent
tests of General Relativity, and explore alternative gravitational 
theories~\cite{Barausse:2012da,Palenzuela:2013hsa,Shibata:2013pra}.
Even more information will be provided by concurrent observations in both gravitational and electromagnetic bands,
potentially establishing direct links between these binaries and spectacularly energetic events 
such as short gamma ray bursts observed already in the electromagnetic spectrum (see e.g.~\cite{Berger:2013jza}).
Additionally, these systems are expected to be prodigious 
producers of neutrinos and, provided
they are sufficiently close, triggers for detectors such as IceCUBE and Super-Kamiokande~\cite{Spiering:2014jna}.
The development of the latest generation of detectors across these channels
promises exciting insights from multi-messenger astronomy.

Extracting such insight from observation, however, requires theoretical predictions
of these very complicated systems.
And this complexity generally requires numerical simulation of a fluid coupled to
relativistic gravity. Moving beyond the most simplified of fluids, those described by
a polytropic EoS, further realism
has been steadily achieved with an ideal gas EoS and, more recently, the adoption of generalized equations of 
state (e.g.~\cite{Duez:2009yy,Hotokezaka:2011dh}). Studies have also explored
the variation of physical parameters such as masses, mass ratios, 
and spin orientations and magnitudes (see~\cite{Etienne:2008re,Chawla:2010sw,Foucart:2010eq,Rezzolla:2010fd}).
Consideration of magnetic fields in these binaries has been pursued with either
ideal magnetohydrodynamics~(MHD)~\cite{Anderson:2008zp,Liu:2008xy,Giacomazzo:2009mp,Chawla:2010sw,Etienne:2012te}
or resistive MHD~\cite{Palenzuela:2013hu}, and incipient steps have
considered neutrino production during the
merger~\cite{Sekiguchi:2012uc,Kiuchi:2012mk,Sekiguchi:2011zd,Deaton:2013sla} or in related
single-star systems~\cite{Kaplan:2013wra,Galeazzi:2013mia} (see also~\cite{Paschalidis:2012ff}).

The effects of magnetic field and neutrino production and cooling
play a sub-leading role in the dynamics of the binary during its
orbiting stages. On the other hand, they can profoundly affect the outcome of the merger and
its subsequent evolution.
In the particular case of a binary neutron star system, to leading order, the lifetime of the 
remnant is determined by the individual stellar masses,
the properties of the EoS, the presence of angular momentum transport mechanisms and cooling effects.
If the total mass of the binary exceeds $\approx 2.6-2.8 M_{\odot}$, prompt collapse to
black hole is expected~\cite{Sekiguchi:2012uc}. Otherwise a hyper-massive neutron star~(HMNS)
is formed, supported by thermal pressure, differential rotation, and the stiffness
provided by its EoS. 
In particular, during the merger kinetic energy is transformed into thermal energy resulting in a very
hot central region (many tens of MeV) providing significant thermal pressure. The angular momentum
given to the remnant from the merger provides for strong, differential rotation
centrifugally
supporting the concentrated mass. Thus, the rates
of cooling and angular momentum transport can have a decisive impact on the onset of delayed black hole formation.

The timescale associated with the transport of angular momentum due to magnetic field
winding is of the order $\tau_{\rm wind} \simeq R/v_{A}$ with $R$ the HMNS radius and
$v_A\simeq B/\sqrt{\rho}$ the \Alfven\,  velocity.
Simulations have found that during the merger the magnetic field strength $B$ can increase
as high as $10^{15-16}$G.
The increase arises from the  compression and winding of the magnetic field  and the
transfer of hydrodynamical kinetic energy to electromagnetic energy 
via Kevin-Helmholtz instabilities and turbulent 
amplification~\cite{Price:2006fi,Anderson:2008zp,Giacomazzo:2010bx,2010A&A...515A..30O,2013ApJ...769L..29Z}. 
For typical densities, these values lead to predicted timescales of 
$\tau_{\rm wind} \simeq 10-100$ms. We note that 
the computational demands of both evolving for this time
period and exploring the complete parameter space make such
computations currently impractical.
Simulations of the magneto-rotational instability, another factor resulting in the transport of
angular momentum, are even more challenging. Its role in non-vacuum, compact object merger
is still largely unexplored.  Nevertheless estimates
indicate a timescale $\tau_{\rm MRI} \approx  100$ms for magnetic field strengths of $B \approx 10^{15}$G. 
Therefore, either effect can operate within timescales $\gtrsim 10 - 100$ms and contribute to the collapse.
In contrast, cooling (via radiation transport) is estimated to act on the order of seconds, and so
while relatively unlikely to impact strongly the dynamics of the HMNS (except in sufficiently low mass binaries), 
the extreme temperatures achieved by the merger can induce a strong neutrino luminosity.

Here we describe our incorporation both of generalized EoSs and of neutrino cooling.
The former is achieved by extending our MHD implementation to handle tabulated equations of state,
and the latter requires adding a suitable approximation to the relevant microphysics. 
A complete treatment of the  microphysics would generally require solving the full transport problem
in three (spatial) dimensions---along with three additional dimensions for the momentum phase space---which
is currently 
out of reach for current (and near-future) computational
resources.
There are various approaches to approximating transport such 
as the truncated moment formalism for radiative hydrodynamics~\cite{Takahashi:2012be,Shibata:2012zz}
and similar work utilizing a radiative transfer code applied to the ejecta 
predicted by their relativistic code to produce light curves for comparison to a recent kilonova prospect~\cite{Hotokezaka:2013kza}.
Fortunately however, the late stages of compact binary mergers occur on short timescales (with the possible
exception of, yet to be observed, low mass binaries) and the details of
radiation transport are subleading with respect to the bulk dynamics of the system.
It therefore suffices to account for the changes to energy and lepton number only locally
in a {\em leakage} approach~\cite{Ruffert:1995fs,Rosswog:2003rv}.

Within efforts to simulate core-collapse supernovae, codes implementing the leakage approach have been constructed, 
validated and distributed publicly~\cite{O'Connor:2009vw,Liebendoerfer:2007dz},
 and we  modify one of these~\cite{O'Connor:2009vw} for use within
our compact binary simulations. 
  In this scheme, the neutrino energy and lepton number
  emission from microphysical processes is determined via an
  interpolation between two limiting regimes: the diffusion limit and
  the free streaming limit. We maintain consistency by accounting for
  the energy and lepton number emission in the fluid variables.
We study the effects of neutrino cooling and compare to previous
 work---where such work exists---for both
single stars and binaries. The results indicate a robust, convergent code consistent with past work.

In~\autoref{sec:implementation}, we describe our implementation of realistic EoS and of the leakage scheme,
and in~\autoref{sec:numericalresults} we present the results of a number of tests. We conclude
in~\autoref{sec:conclusions}.

\section{Implementation}
\label{sec:implementation}

Our previous studies on neutron stars used a perfect fluid with
an ideal equation of state with $\Gamma=2$. In this paper we detail only 
the recent changes to our code to use finite temperature equations of state
and the leakage scheme. However, we also briefly present the Einstein and 
fluid equations to define our notation. 

\subsection{Evolution equations}
From a physical point of view, we follow the dynamics of both gravitational
and magnetohydrodynamical fields. The former is unchanged with respect to our
previously described work~\cite{Neilsen:2010ax}. However, the MHD
equations
(e.g.~\cite{Neilsen:2005rq,Liebling:2010qv})
 must be modified to account for the effect of neutrinos in both energy
and lepton numbers. Since the leakage scheme is essentially a local calculation (the
exception being the optical depth)
providing lepton and energy rates of change as measured by a co-moving observer,
its extension to the general relativistic case is straightforward. 
For the sake of completeness we describe next the basic strategy.

The Einstein equations in the presence of both matter and radiation are
\begin{equation}
G_{ab} = 8 \pi (T_{ab} + {\cal R}_{ab} )
\label{Eins1}
\end{equation}
where $T_{ab}$ is the stress energy tensor of a perfect
fluid, ${\cal R}_{ab}$ is the contribution from the radiation field, and
we have adopted geometrized units where $G=c=M_\odot = 1$. 
Eq.~\ref{Eins1}, coupled with appropriate prescriptions for the dynamics 
of $T_{ab}$ and  ${\cal R}_{ab}$, defines the system of equations. In what
follows we briefly describe how  each is implemented.

We solve the Einstein equations by adopting a 3+1 decomposition in terms
of a spacelike foliation. The hypersurfaces that constitute this
foliation are labeled by a time coordinate $t$ with unit normal $n^a$ and
endowed with spatial coordinates $x^i$. We express the spacetime metric as
\begin{equation}
ds^2 = -\alpha^2\,dt^2 
+ \gamma_{ij}\left(dx^i + \beta^i\,dt\right)\left(dx^j + \beta^j\,dt\right),
\end{equation}
with $\alpha$ the lapse function and $\beta^i$ the shift vector.
Specifically,
we express  the Einstein equations in terms of the BSSN-NOK 
formalism~\cite{1987PThPS..90....1N,1995PhRvD..52.5428S,1999PhRvD..59b4007B,
lousto}.
In this formulation, the metric on spatial hypersurfaces,
$\gamma_{ij}$, is expressed in terms of a conformal factor $\chi$
and a conformally flat metric $\tilde \gamma_{ij}$
\begin{equation}
\gamma_{ij} = \frac{1}{\chi}\tilde\gamma_{ij}, \qquad 
         \chi = \left(\det \gamma_{ij}\right)^{-1/3}
\end{equation}
such that $\det\tilde\gamma_{ij}=1$. In addition,
the extrinsic curvature $K_{ij}$ is decomposed into its trace
$K\equiv K^{i}{}_i$ and the conformal, trace-less extrinsic curvature
$
\tilde A_{ij} = \chi\left(K_{ij} - \frac 1 3 \gamma_{ij}K\right).
$
Finally, one introduces the conformal connection functions
$
\tilde\Gamma^i = \tilde\gamma^{jk}\tilde\Gamma^i{}_{jk}
$
which are evolved as independent variables.
The evolution equations are
\begin{align}
\partial_t {\tilde \gamma}_{ij} 
    & = \beta^k \partial_k {\tilde \gamma}_{ij} 
+ {\tilde \gamma}_{ik} \, \partial_j \beta^k 
+ {\tilde \gamma}_{kj} \partial_i \beta^k  \nonumber \\
& - {2\over3} \, {\tilde \gamma}_{ij} \partial_k \beta^k 
- 2 \alpha {\tilde A}_{ij} \\ 
\partial_t\chi &= \beta^i\partial_i \chi +
  \frac{2}{3}\chi\left(\alpha K -\partial_j\beta^j\right) \\
\partial_t \tilde A_{ij} &= 
  \beta^k \partial_k{\tilde A}_{ij} 
+ {\tilde A}_{ik} \partial_j \beta^k 
+ {\tilde A}_{kj} \partial_i \beta^k 
- {2\over3} \, {\tilde A}_{ij} \partial_k \beta^k \nonumber \\
&+ \chi\left[-D_i D_j\alpha 
       + \alpha\left(R_{ij} - 8\pi S_{ij}\right)\right]^{{\rm TF}}\nonumber\\
       &+ \alpha\left(K\tilde A_{ij} - 2\tilde A_{ik} \tilde A^k{}_j\right)\\
\partial_t K &= \beta^k \partial_k K  -D^iD_i\alpha\nonumber\\
   &+ \alpha\left[\tilde A_{ij}\tilde A^{ij} + \frac{1}{3}K^2 
    + 4\pi\left(E+S\right)\right] \\
\partial_t \tilde \Gamma^i &= 
\beta^j \partial_j {\tilde \Gamma}^i
- {\tilde \Gamma}^j \partial_j \beta^i
+ {2\over3} {\tilde \Gamma}^i \partial_j \beta^j 
+ {\tilde \gamma}^{jk} \partial_j \partial_k \beta^i \nonumber\\
&+ {1\over3} \, {\tilde \gamma}^{ij} \partial_j \partial_k \beta^k
- 2\tilde A^{ij}\partial_j\alpha \nonumber \\
&+ 2\alpha\left(\tilde\Gamma^i{}_{jk}\tilde A^{jk}
- \frac{3}{2\chi}\tilde A^{ij}\partial_j \chi
-\frac{2}{3}\tilde\gamma^{ij}\partial_j K 
- 8 \pi {\tilde \gamma}^{ij}S_j\right).
\end{align}
In these equations, the matter terms are defined as
\begin{align}
E &\equiv n_a n_b \left( T^{ab} + {\cal R}^{ab} \right )\\
S_i &\equiv -\gamma_{ia} n_b \left( T^{ab}+ {\cal R}^{ab} \right )\\
S_{ij} &\equiv \gamma_{ia} \gamma_{jb} \left( T^{ab}+ {\cal R}^{ab} \right ).
\end{align}
Notice that for the problem of interest $|{\cal R}^{ab}| \ll
|T^{ab}|$. This observation, together with the fact that
a neutrino leakage scheme can not possibly treat the radiation
stress tensor fully consistently, we choose to ignore the source-term-contribution from
the radiation field to the Einstein equations. 

The evolution equations
are supplemented with gauge conditions. We use the ``1+log'' slicing
condition and the $\Gamma$-driver shift with the evolution equations
\begin{align}
\partial_t\alpha &= \lambda_1 \beta^i\partial_i\alpha - 2\alpha K\\
\partial_t \beta^i &= \lambda_2 \beta^j\partial_j\beta^i 
       + \frac{3}{4}f(\alpha)B^i\\
\partial_t B^i &= \partial_t\tilde\Gamma^i - \eta B^i 
      + \lambda_3 \beta^j\partial_j B^i 
      - \lambda_4\beta^j\partial_j\tilde \Gamma^i.
\end{align}
Here $f(\alpha)$ is an arbitrary function and 
$\lambda_1$, $\lambda_2$, $\lambda_3$, $\lambda_4$, and $\eta$
are parameters that can be chosen for different types of initial data. 
Our simulations are performed with the choice 
$f(\alpha) = \lambda_i = 1$ and $\eta \approx 3.5/M$, 
where $M$ is the total mass of the system.
Finally, during the evolution the algebraic constraints
\begin{equation}
{\rm det}\tilde\gamma_{ij} = 1, \qquad \tilde A^{i}{}_i = 0
\end{equation}
are enforced at every step.

\subsection{Matter}

For the matter source we consider a perfect fluid with stress energy tensor given by
\begin{equation}
T_{ab} = h u_a u_b + P g_{ab}  
+ F_{ac} F^c_b - \frac{1}{4} g_{ab} F_{cd} F^{cd}\, .
\end{equation}
where $h$ is the {\it total} 
enthalpy $h =  \rho (1+\epsilon) + P$,
and $\{\rho,\epsilon,u^a,P\}$ are the rest mass energy density, 
specific internal energy,
four-velocity and pressure respectively, and $F_{ab}$ the Faraday tensor
(absorbing a factor $1/\sqrt{4 \pi}$ in its definition).
Provided an equation  of state of the form $P=P(\rho,\epsilon,Y_e)$ 
and a relativistic Ohm's law,
the equations determining the magnetized
matter dynamics are obtained from suitable conservation laws. We here adopt
the ideal MHD approximation (which states that the fluid is described by
an isotropic Ohm's law with perfect conductivity, so that the electric
field vanishes in the fluid's frame $F_{ab} u^b =0$)
which provides a simple, realistic approach reducing the number of relevant
fields to describe electromagnetic effects in the system.

The resulting system of equations is
\begin{align}
\nabla_a T^a_b &= {\cal G}_b \label{eq:DT}\\
\nabla^a (T_{ab} n^b) &= 0 \label{eq:DTN} \\
\nabla_a (Y_e \rho u^a) &= \rho R_Y \label{eq:divYe}\\
\nabla_a {}^* F^{ab} &= 0 \label{eq:DF}.
\end{align}
These equations state
the conservation laws for the stress-energy tensor, matter and
lepton number respectively where  $Y_e$ is the electron fraction, the
ratio of electrons to baryons. In the absence of lepton source terms,
Eq.~(\ref{eq:divYe}) follows closely the conservation law for the
rest mass density, i.e. $Y_e$ is a mass scalar.  The sources ${\cal
  G}_a$~($\equiv -\nabla_c {\cal R}^c_{a}$) and $R_Y$ are the radiation
four-force density and lepton sources which are determined via the
leakage scheme.

\subsubsection{GR-Hydro equations}
Examining the equations for the matter, Eqs.~(\ref{eq:DT}-\ref{eq:DF}),
a few observations are in order.
On the one hand, the implementation of the equations for the magnetic field need not be modified
with respect to our previous work (e.g.~\cite{Liebling:2010bn}) as they are neither directly 
coupled to the neutrino evolution nor do they depend on the fluid's equation of state.
On the other hand, changes to the implementation of the fluid equations are required. In what follows
we describe those modifications.

Recall the 3+1 decomposition of the  fluid equations 
as presented in~\cite{Neilsen:2005rq}.
The relevant expressions of the projections with respect to $n^a$ (parallel and orthogonal)
can be written in terms of the source  ${\cal G}_a$ as
\begin{eqnarray}
0 &=& -n^a \partial_a E + KE - \frac{1}{\alpha^2}D_a(\alpha^2 S^a) + \nonumber \\
& & (\perp T)^{ab} K_{ab} - {\cal G}_a n^a \\
0 &=& h_{bc} \left [ -n^a \partial_a S^b + K S^b + 2 S^a K_a^b
-\frac{1}{\alpha} S^a \partial_a \beta^b \right. \nonumber \\
& & - \left. \frac{1}{\alpha} D_a\left(\alpha(\perp T)^{ab} \right) 
- \frac{\partial^b \alpha}{\alpha} E + {\cal G}^c \right ].
\end{eqnarray}
Finally, we define the Lorentz factor and the three-velocity as
\begin{equation}
W \equiv -n^a u_a, \qquad v^i \equiv \frac{1}{W}(\perp u)^i.
\end{equation}
The fluid equations of motion are written in balance law form
\begin{equation}
\partial_t {\bf u} + \partial_i{\bf f}^i\left({\bf u}\right) = {\bf s}({\bf u})
\end{equation}
by defining the {\it conservative} variables. These variables are densitized
using the 3-metric determinant $\sqrt{\gamma}$ as
\begin{align}
\tilde D&\equiv \sqrt{\gamma}\rho W\\
\tilde S_i&\equiv \sqrt{\gamma}\left[\left(h W^2 + B^2\right)v_i - \left( B^j v_j\right) B_i\right]\\
\tilde \tau &\equiv \sqrt{\gamma}\left[ h W^2 + B^2 - P - \frac{1}{2}\left(\left(B^i v_i\right)^2 + \frac{B^2}{W^2}\right)\right]\\
\tilde B^i &\equiv \sqrt{\gamma} B^i\\
\tilde {Y_e}  &\equiv \tilde D Y_e.
\end{align}

The leakage scheme provides the
  fluid rest frame energy sink ${\cal Q}$ and lepton sink/source $R_Y$ due to
  neutrino processes.  $R_Y$ is the source term for a scalar quantity
  and therefore is the same in all frames.  We express the source term
  for the energy and momentum in an arbitrary frame as
\begin{equation}
{\cal G}_{a} = {\cal Q} u_a \, .
\end{equation}
Since the effect
  of neutrino pressure is small~\cite{Deaton:2013sla} and difficult
  to accurately capture with a neutrino leakage scheme, we ignore its
  contribution in the fluid rest frame.  Now, defining ${\cal H}
\equiv n^a {\cal G}_a = - {\cal Q} W $ and ${\cal H}_b \equiv h_{bc}
{\cal G}^c = {\cal Q} W v_b$, the modified relativistic
MHD equations become
\begin{align} 
&\partial_t\tilde D + \partial_i \left[ \alpha\tilde D \left( v^i - {\beta^i \over \alpha} \right) \right] = 0 \\
&\partial_t \tilde S_j 
   + \partial_i\left[\sqrt{-g}\left(\left(\perp\! T\right)^i_j 
   - \frac{\beta^i}{\alpha} S_j\right)\right]\nonumber\\
&\quad =\sqrt{-g}\,\left[\,{^{3}{\Gamma}}_{ab}^i\left(\perp\! T\right)^a{}_i 
    + {1\over\alpha} S_a \partial_b \beta^a 
    - {1\over\alpha} \partial_b \alpha \, E  + {\cal H}_b\right] \\
&\partial_t \tilde \tau + \partial_i \left[ \alpha \left( \tilde S^i - \tilde D v^i - {\beta^i \over \alpha} \, \tilde \tau \right) \right] \nonumber \\
&\qquad =\sqrt{-g} \, \left[ \left( \perp\! T \right)^{ab} \, K_{ab} - {1\over\alpha} \, S^a \partial_a \alpha - {\cal H} \right] \\
&\partial_t \tilde{Y_e} + \partial_i\left[\alpha\tilde{Y_e}
      \left(v^i-\frac{\beta^i}{\alpha}\right)\right] 
      = \frac{\alpha}{W}\tilde D R_Y\\
&\partial_t {\tilde B}^i +
   \partial_j \left[  \left( v^j - \frac{\beta^j}{\alpha}\right) {\tilde B}^i
                    - \left( v^i - \frac{\beta^i}{\alpha}\right) {\tilde B}^j  \right]
    \nonumber \\
&\qquad = - \alpha \sqrt{\gamma} \gamma^{ij} \partial_j \psi \\
&\partial_t \psi =  - c_h^2 \frac{\alpha}{\sqrt{\gamma}}\partial_i {\tilde B}^i
    - \alpha c^2_r \psi
\end{align}
where
\begin{align}
(\perp T)^i{}_j &= v^iS_j + P h^i{}_j - \frac{1}{W^2}\left[ B^i B_j - \frac{1}{2}h^i_j B^2\right]\nonumber\\
& - \left(B^kv_k\right)\left[B^i v_j - \frac{1}{2} h^i{}_j B^m v_m\right].
\end{align}

Notice that the evolution equation for ${\tilde B}^i$ contains a scalar field ($\psi$)
contribution which is introduced to control the no-monopole constraint.
This ``divergence cleaning method'' damps constraint violations via
a damped, wave equation for the evolution of  $\psi$~\cite{Dedner2002,Neilsen:2005rq}.

Notice also that this system of equations is strongly hyperbolic. It contains the 
minimum couplings with the solenoidal constraint and includes a damping term.
The terms in the right hand sides are treated as sources, and their derivatives are calculated
with centered finite difference (second order).
A small amount of numerical dissipation is added both to $\{ {\tilde B}^i , \psi \}$.
We have found numerically that an efficacious choice is $c_h=0.1$, $c_r \approx 2/\sqrt{M}$
(where $M$ is again the total mass of the system).

\subsection{Equations of State}

\subsubsection{Finite temperature EoS tables}

Neutrino interaction rates depend sensitively on the matter
temperature and composition. Therefore, in order to model the
effect of neutrinos with reasonable accuracy, we require an equation of state beyond
that of a polytrope or an ideal gas. We use publicly available
EoS tables from www.stellarcollapse.org and described in
O'Connor and Ott~(2010)~\cite{O'Connor:2009vw}.  We have rewritten
some of the library routines for searching the table to make them
faster and more robust. In this paper we use the 
Shen-Horowitz-Teige~(SHT)~\cite{2011PhRvC..83c5802S} EoS with the NL3
relativistic mean-field parametrization, the 
Lattimer-Swesty~(LS)~\cite{1991NuPhA.535..331L} EoS with $K=220$ MeV,
and the H.~Shen~(HS)~\cite{2011ApJS..197...20S} for the single neutron
star simulations, and the HS EoS for the neutron star binary.

\subsubsection{The primitive solver}

High-resolution shock-capturing schemes integrate the fluid equations
in conservation form for the conservative variables, while the fluid
equations are written in a mixture of conserved and primitive
variables. It is well known that the 
calculation of primitive variables from conserved variables
for relativistic fluids  
requires solving a transcendental set of equations. Our method for
solving these equations with a finite-temperature
EoS is a modification of the algorithm that we use
for the ideal gas EoS;
the most significant change being that the internal energy
must be calculated separately from the pressure using the table. 
We write the transcendental equations in terms of the new variable
\begin{equation}
x \equiv h W^2,
\end{equation}
where  $h$ is the {\em total} enthalpy $h = \rho(1+ \epsilon) + P$,
and calculate $Y_e$ from the evolution variables  $\tilde Y_e/\tilde D$.
Then, using data from the previous time step to calculate an initial guess
for $x$, we iteratively solve these equations for $x$:
\begin{enumerate}
\item From the equation for $S^iS_i$, calculate
\[
\hat W^{-2} =
1 - \frac{(2x + B^2)B^iS_i + x^2S^2}{\left( x \left(x + B^2\right)\right)^2}.
\]
\item From the definition of $D$, calculate
\[\hat \rho = \frac{D}{\hat W}.\]

\item  From the definition of $\tau$, calculate
\[\hat P = x - (\tau + D) + B^2 
- \frac{1}{2}\left[\left(\frac{B^iS_i}{x}\right)^2 + \frac{B^2}{{\hat W}^2}\right].\]

\item From the definition of the total enthalpy, calculate
\[
\hat\epsilon=\left(\frac{x}{\hat W^2}-\hat P\right)\frac{\hat W}{D}-1.
\]

\item Use the EoS table to calculate the temperature $\hat T$ from 
$\hat \epsilon(\hat\rho, \hat T, Y_e)$.

\item Use the EoS table to calculate $P(\hat\rho, \hat T, Y_e)$.

\item Update the guess for $x$ by solving the equation $f(x)=0$ using
the Brent method, with (again, the definition of $\tau$)
\[f(x) = x - P(\hat\rho, \hat T, Y_e) 
- \frac{1}{2}\left[\left(\frac{B^iS_i}{x}\right)^2 + \frac{B^2}{{\hat W}^2}\right]
+ B^2 - (\tau + D). \]
\end{enumerate}
The root of $f(x)=0$ from Step~7 becomes the new guess for $x$, and this
process is repeated iteratively until the solution for $x$ converges 
to a specified tolerance. One advantage of this algorithm is that $f(x)$ is
a function of a single variable, and,
in contrast to root solving for multiple variables, robust methods
can be used to find any root that can be bracketed.

Because of numerical error, a solution to these equations may either fall
outside the physical range for the primitive variables, or a real solution
for $x$
may not exist. The solutions for  $\rho$, $T$, and $Y_e$
are, at a minimum, restricted to values in the table,
and they are reset to new values (the minimum allowed value plus ten percent) 
if necessary.  
A separate floor value for the density is also set. In anticipation of
comparing this work with evolutions of magnetized stars in the future, we choose
a density floor appropriate for magnetized stars for the neutron star binary,
which is about eight orders of magnitude 
smaller than the initial central density of the stars.
If a real solution for the 
primitive variables does not exist, the primitive variables are
interpolated from neighboring points, and the conserved variables are reset
to be consistent. If a valid interpolation stencil can not 
be constructed because the solver also failed at the neighboring points,
then the update fails, and the run is terminated. 
This failure occurs very rarely and 
may be remedied by slightly increasing the density floor.

\subsection{Leakage}

The leakage scheme seeks to account for (i) the changes to the
(electron) lepton number and (ii) the loss of energy from the emission
of neutrinos.  As discussed, since the dynamical timescale for the
post-merger of binary neutron star systems is relatively short,
radiation momentum transport and diffusion effects are expected to be
subleading.  Our scheme is based on the open-source neutrino leakage
scheme from \cite{O'Connor:2009vw} and available at
www.stellarcollapse.org.  At low optical depths, the leakage scheme
relies on calculating the emission rate of energy ($Q_\mathrm{free}$)
and lepton number ($R_\mathrm{free}$) directly from the rates of
relevant processes.
We consider three species of neutrinos, represented here by:
$\nu_e$ for electron neutrinos, $\bar{\nu_e}$ for electron antineutrinos,
and $\nu_x$ for both tau and muon neutrinos and their respective antineutrinos.
As discussed in~\cite{Ruffert:1995fs,Rosswog:2003rv}, the dominant processes are
those that
 \begin{itemize}
               \item{produce electron flavour neutrinos and antineutrinos: charged-current, electron and positron capture reactions\\
               $e^++n\rightarrow p + \bar{\nu}_e$ ~, ~
               $e^-+p\rightarrow n +     {\nu}_e$ ~.}
               \item{produce all flavours of neutrinos: electron-positron pair-annihilation\\
               $e^++e^- \rightarrow \bar{\nu}_i + \nu_i$ ~~~
            
               and plasmon decay\\
               $\gamma \rightarrow \bar{\nu}_i + \nu_i$.
               }
 \end{itemize}
Notice that nucleon-nucleon bremsstrahlung can also be an important source of
$\nu_x$ neutrinos, dominating over electron-positron annihilation
at low temperatures and high densities. We will include such a process
in future work.

At high optical depths on the other hand, because the equilibrium time
scales are much shorter than either neutrino diffusion or hydrodynamic
time scales, neutrinos are assumed to be at their equilibrium
abundances and the rates of energy loss ($Q_\mathrm{diff}$) and lepton
loss ($R_\mathrm{diff}$) are taken to proceed at the diffusion
timescale. The equilibrium abundances can be trivially calculated,
however the calculation of the diffusion timescale is more involved as
it requires the knowledge of non-local optical depths.  The
computation of these optical depths lies at the core of the leakage
strategy and, because we are interested in general (non-spherically
symmetric) scenarios, we describe how to compute them from the local
opacities in~\autoref{sec:optdepth}. (We refer the reader to
\cite{O'Connor:2009vw} for full details about the calculation of the local opacity and
diffusion time scale.)
The emission rates are then interpolated between the behavior at low
and high optical depths in order to achieve an efficient way to
incorporate neutrino effects that is correct in both regimes and
applicable in between. In our implementation, we interpolate the
energy and lepton number emission rates between these two regimes via
the following formula
\begin{equation}
X_\mathrm{eff} = \frac{X_\mathrm{diff}X_\mathrm{free}}{X_\mathrm{diff}
+ X_\mathrm{free}}\,,
\end{equation}
where $X$ is either $Q$ or $R$.

\subsubsection{Optical depth calculation}
\label{sec:optdepth}

\begin{figure}[h]
\centering
\includegraphics[width=8.5cm,angle=0]{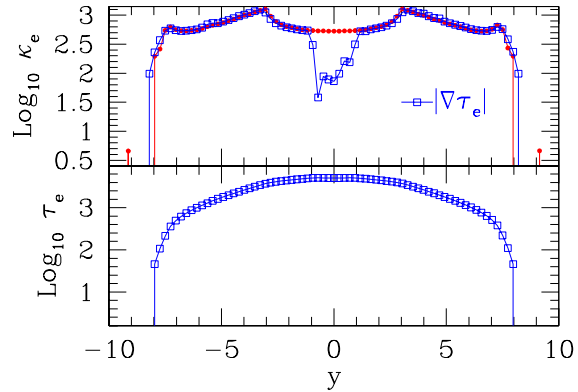}
\vspace{-0.85in}
\caption{{\em Optical depth for an isolated neutron star}.
{\bf Top:} The electron opacity $\kappa_e$ (red solid) evolved for $t=1.15ms$ along the $x$-axis. 
Also shown is the magnitude of the gradient of the electron optical depth $|\nabla \tau_e|$.
The noticeable dip  in the gradient is just an effect of the post-processing used to compute
the gradient and no significant feature is seen in the optical depth itself shown in the
bottom panel.
{\bf Bottom:}  The electron optical  depth $\tau_e$.
}
\label{fig:optdepthNS}
\end{figure}
\begin{figure}[h]
\centering
\includegraphics[width=8.5cm,angle=0]{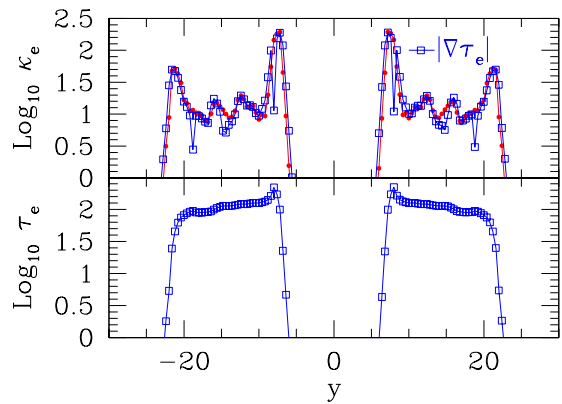}
\vspace{-0.85in}
\caption{{\em Optical depth for a binary neutron star system}.
{\bf Top:} The electron opacity $\kappa_e$ (red solid) after a quarter orbit along the $y$-axis. 
Also shown is the magnitude of the gradient of the electron optical depth $|\nabla \tau_e|$.
The similarity of these two quantities suggests that the optical depth algorithm adequately tracks the binary members.
{\bf Bottom:}  The electron optical  depth $\tau_e$. The binary is quite cold (initial temperature of $0.01$ MeV) and so the depth is quite small.
}
\label{fig:optdepthBNS}
\end{figure}

The usual approach to calculating the optical depth at a given point is to consider some small number of possible directions in which to integrate the opacity
of the fluid. In {\tt GR1D}~\cite{O'Connor:2009vw}, the assumption of spherical symmetry simplifies the calculation so that there are only ingoing and outgoing directions.

Refs.~\cite{Sekiguchi:2012uc} and~\cite{Deaton:2013sla} integrate the
depth along rays in the coordinate directions, although Ref.~\cite{Deaton:2013sla} adds certain diagonal rays. The depth at any given point is then the
minimum depth among the considered rays.
Ref.~\cite{Galeazzi:2013mia} instead argues for rays that match the geometry of the problem and they therefore interpolate onto a spherical grid and considers the minimum depth among a set of radial rays.

In general, the existent algorithms necessarily involve global integrations that bring with them complexities due to multiple resolutions (from the AMR) and patches (from the domain decomposition). Instead, a more local approach that is independent of the particular symmetries of the problem is desirable. For example, one could determine the depth at any given point as a parallel circuit where the depth measures the ``resistance'' to the depth-free exterior~\footnote{Suggested by Philip Chang.}. The depth at each point is just the contribution to the depth to reach a neighboring point added to the inverse of the sum of the inverses of the depths of all neighboring points. Such a procedure turns the global integration into an iterated local problem. However, early experiments showed a problem with its application because the depth was generally less than that of any neighbors. Therefore, in a naive iteration scheme, the resulting depth did not adequately reflect the expected increasing optical depth with increased depth into the star.

However, an even simpler approach appears to work quite well. In this scheme, the depth at any given point is simply
the sum of the depth incurred to get to a neighboring point plus the minimum depth among its neighbors.
One can justify such an approach by arguing that neutrinos will explore all pathways out of the star, not just 
straight paths. This approach is also iterative since changes elsewhere do not immediately affect other areas,
as would happen with a global integration. Physically one expects changes at the surface to take some time
to propagate throughout the star. However, as noted in~\cite{O'Connor:2009vw}, because the depth
depends on the opacity which itself depends on the depth, one expects to iterate in any case. 

We have compared runs with no explicit iteration (our leakage routine is called at every Runge-Kutta step, instead of once per time step) with runs where we iterate three times and there is effectively no difference.

The computation of the optical depth appears to be an example of the eikonal problem\footnote{Pointed-out to us by Ian Hawke.}. The eikonal equation takes the form
\begin{equation}
|\nabla u(x)| = f(x)
\end{equation}
for scalar functions $u(x)$ and $f(x)$ (see for example Ref.~\cite{zhao05} for a fast method of solving it).
Here, this equation takes the form
\begin{equation}
|\nabla \tau_i(x)| = \kappa_i(x)
\end{equation}
where $\tau_i$ is the optical depth for some species of neutrino and $\kappa_i$ its corresponding opacity.
Implementation of one of the approaches to the eikonal problem for the computation of optical depth may provide benefits,
but 
here we simply use this formulation (the derivative of the defining integral for the depth) to evaluate the utility of our admittedly simplified solution.

In Fig.~\ref{fig:optdepthNS} we show both the magnitude of the gradient of the depth for electron neutrinos and compare it with its opacity for a single star. The agreement between these two indicates
that even this simplified scheme is finding the correct optical depth for the given opacity.

One possible concern is that this approach is too simple to track stars in a binary and so in Fig.~\ref{fig:optdepthBNS} we do the same analysis along the $y$-axis for a binary when the stars are suitably aligned.
Once again, the features of the gradient match those of the opacity, and, in particular, the depth tracks the stars throughout their orbit.

\section{Numerical results}
\label{sec:numericalresults}
We present initial tests and preliminary results below. A more detailed
and expansive study of binary neutron stars will be presented in a future
paper.

\subsection{Isolated Stars}

A standard test consists of evolving isolated, neutron stars 
and analyzing their oscillation modes. The frequencies of these modes
can be computed independently by solving for the linearized perturbations
in the Cowling approximation (i.e., fixed spacetime). 
The extent to which
the frequencies obtained from our fully non-linear evolution agree
with those of the linearized code helps measure the correctness of our code.

We constructed initial data for our neutron stars using three different
nuclear EoSs describing hot
dense matter, all of them publicly available in~\cite{ott_eos_table}, using
{\sc Magstar}, part of {\sc Lorene}~\cite{lorene}.
The first one corresponds to the stiff EoS by
Shen-Horowitz-Teige~(SHT)~\cite{2011PhRvC..83c5802S} using the NL3
relativistic mean-field parametrization, the second one is by
Lattimer-Swesty~(LS)~\cite{1991NuPhA.535..331L} with an
incompressibility modulus $K=220$ MeV, representing a soft EoS,
and the final one is the H.~Shen~(HS)~\cite{2011ApJS..197...20S} developed from
relativistic mean-field theory with the TM1 parametrization. These equations of state have cold
neutron star maximum gravitational masses of $2.76\,M_\odot$,
$2.04\,M_\odot$, and  $2.24\,M_\odot$, respectively.

The simulations are performed on a numerical domain covering $x^i \in
[-60{\rm km},60{\rm km}]$ for the stars evolved in a fixed background,
and it extends to $x^i \in [-120{\rm km},120{\rm km}]$ when the
spacetime is dynamical in order to prevent unphysical effects coming
from the boundary. There are several refinement levels so that there
is a grid covering the star with a resolution $\Delta x = 250 {\rm m}$.
We evolve the system with a third order accurate, Runge-Kutta scheme with a time
step given by $\Delta t = 0.25 \Delta x $, satisfying the CFL condition.

\begin{table*}
 \begin{tabular}{|l|ccccc|cc|cc|cc|}
\hline
\hline
Model & Metric & ~~$ M{\rm [M_{\odot}]}$ & ~~$ R[{\rm km}]$ & ~~$ s{\rm [k_{B}]}$ & ~~$B_c$ [G] 
      & ~~$f_1$ [KHz] & ~~$f_1^{C}$ [KHz] & ~~$f_2$ [KHz] & ~~$f_2^{C}$ [KHz] & ~~$f_3$ [KHz] & ~~$f_3^{C}$ [KHz] \\
\hline
SHT     & fixed    & ~~$ 2.73 $ & ~~$ 13.90 $ & ~~$ 0 $ & ~~$0$  & ~~$3.54$ & ~~$3.49$ & ~~$5.87$  & ~~$5.86$ & ~~$8.34$  & ~~$8.24$ \\
LS-K220 & fixed    & ~~$ 1.69 $ & ~~$ 12.36 $ & ~~$ 0 $ & ~~$0$  & ~~$3.96$ & ~~$3.89$ & ~~$6.84$  & ~~$6.81$ & ~~$9.95$  & ~~$9.72$ \\
LS-K220 & dynamic  & ~~$ 1.69 $ & ~~$ 12.36 $ & ~~$ 0 $ & ~~$5\times10^{14}$  & ~~$2.37$ & ~~$2.39$ & ~~$6.07$  & ~~$6.09$ & ~~$9.23$  & ~~$9.41$ \\
SHT     & fixed    & ~~$ 2.74 $ & ~~$ 14.14 $ & ~~$ 1 $ & ~~$5\times10^{14}$  & ~~$3.53$ & ~~$3.42$ & ~~$5.90$  & ~~$5.73$ & ~~$8.32$  & ~~$8.01$ \\
\hline
\hline
\end{tabular}
\caption{Mode frequencies of the oscillations of a NS with $\rho_c = 9.3\times10^{14}
{\rm g/cm^3}$. We compute the
first three frequencies (i.e., the fundamental radial mode and the two first
overtones) $f_i$ and compare with frequencies $f^C_i$ from either a linearized
code (fixed spacetime) or from another non-linear code~\cite{Galeazzi:2013mia}
(dynamic spacetime). Note that in the last two cases the
star contains a poloidal magnetic field and the oscillation of the central
magnetic field matches that of the fluid density (see the bottom panels of
Figs.~\ref{fig:rhoc_lsk220_dynamical} and~\ref{fig:rhoBLc_gshen_hot_cowling}).
}
\label{tab:stars}
\end{table*}

\subsubsection{Cold stars}

\begin{figure}[h]
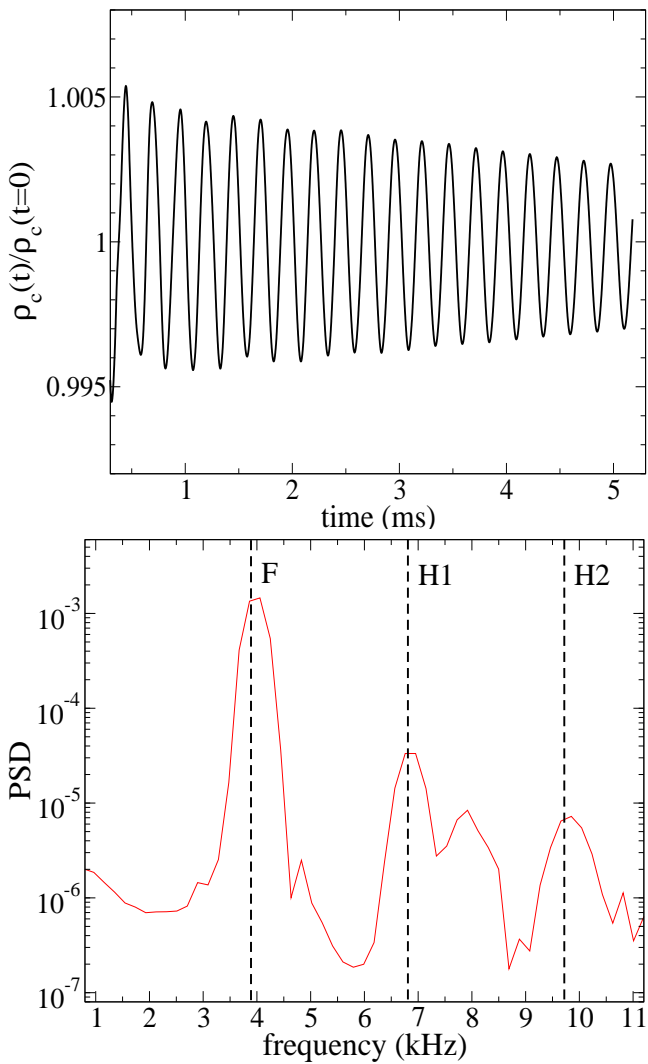

\centering
\includegraphics[height=7.0cm,width=8.5cm,angle=0]{./figures/rhoc_lsk220_cowling.eps}\\
\includegraphics[height=7.0cm,width=8.5cm,angle=0]{./figures/fourier_rhoc_lsk220_cowling.eps}
\caption{{\em Perturbed NS with the  LS-K220 EoS in the Cowling approximation}. {\bf Top:} Central density of the
star as a function of time.
{\bf Bottom:}  The Fourier power spectral density as a function of frequency. Vertical dashed lines show the expected
oscillation frequencies calculated by solving the linearized equations.}
\label{fig:rhoc_lsk220_cowling}
\end{figure}

\begin{figure}[h]
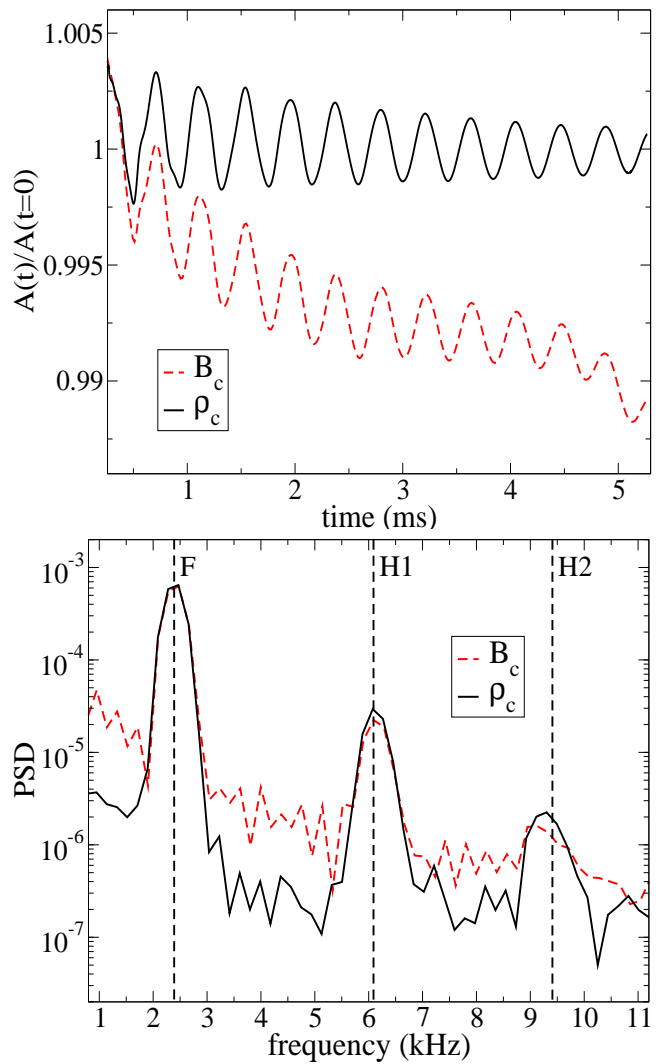

\centering
\includegraphics[height=7.0cm,width=8.5cm,angle=0]{./figures/rhoBc_lsk220_dynamical.eps}\\
\includegraphics[height=7.0cm,width=8.5cm,angle=0]{./figures/fourier_rhoBc_lsk220_dynamical.eps}
\caption{{\em Perturbed, magnetized NS with the  LS-K220 EoS in a dynamic spacetime}. {\bf Top:} The (normalized) central density of the star
and central magnetic field strength as a function of time.
The initial star has a purely poloidal magnetic field.
{\bf Bottom:} The Fourier power spectral density as a function of frequency.
Vertical dashed lines show the computed oscillation frequencies
obtained by using another full non-linear evolution code.}
\label{fig:rhoc_lsk220_dynamical}
\end{figure}

The cold star initial data are constructed assuming a very low
temperature of $T=0.01$ MeV---well below the Fermi temperature
of the star---and imposing $\beta$-equilibrium. An initial
perturbation is introduced by increasing  the
initial temperature with respect to the temperature assumed during its construction to $T=0.05$\,MeV. We
choose a solution on the stable branch of non-rotating stars---near
the maximum allowed mass---to coincide with that of
Ref.~\cite{Galeazzi:2013mia} to facilitate comparison, corresponding
to a central density $\rho_c = 9.3\times10^{14} {\rm g/cm^3}$.

We work in the Cowling approximation such that the metric is frozen
at its initial profile so that we can compare easily with the results of
perturbations from the linearized system. In particular, we Fourier
transform the time-series data for the central density
from the evolution code as shown in Fig.~\ref{fig:rhoc_lsk220_cowling}
for the star using the LS EoS. The results from linear analysis are shown
as dashed lines, and these frequencies correspond quite well to the
peaks of the Fourier power spectrum. 
The numerical values for the first three frequency modes $f_i$ (i.e., the
fundamental radial mode and the first two overtones) are summarized
in Table~\ref{tab:stars} and are compared with the 
frequencies $f_i^C$ computed either from perturbation theory (for fixed
spacetimes) or with another full non-linear code (for the dynamical
spacetimes considered below). The maximum disagreement between these values is
$2.5\%$, and generally below $1.5\%$. Note that, as happens with cold
stars described by a polytropic EoS (see for instance \cite{2002PhRvD..65h4024F}),
the mode frequencies are higher in fixed spacetimes than
in dynamical ones.

With this same star, we can add an initial seed magnetic field. Here
we add a poloidal field with a maximum strength of
$8 \times 10^{14} {\rm G}$, and evolve
the full spacetime.
In this case we transform
the central value of both the density and the magnetic field,
as shown in Fig.~\ref{fig:rhoc_lsk220_dynamical}.
Both spectra lead to the same frequencies. Because there are no published
results from perturbation theory for the dynamical spacetime case, we are
only able to compare the observed oscillation frequencies
with the ones obtained from another independent, fully non-linear code,
Ref.~\cite{Galeazzi:2013mia}, which also solves the general relativistic
hydrodynamic equations with a neutrino leakage scheme.

\subsubsection{Hot stars}

\begin{figure}[h]
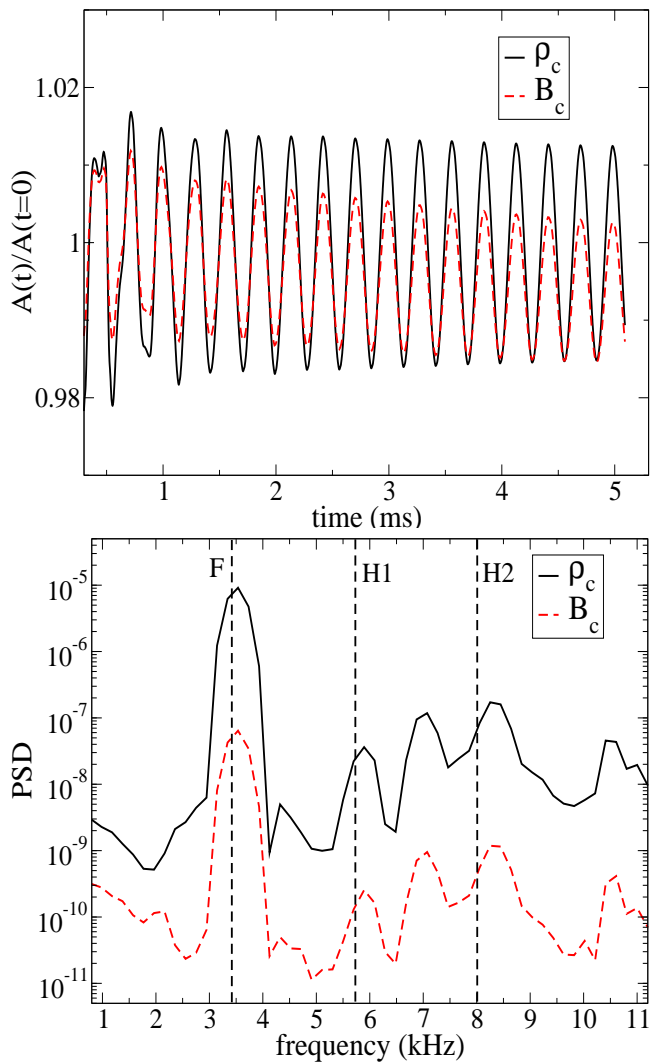

\centering
\includegraphics[height=7.0cm,width=8.5cm,angle=0]{./figures/rhoBc_gshen_hot_cowling.eps}\\
\includegraphics[height=7.0cm,width=8.5cm,angle=0]{./figures/fourier_rhoBLc_gshen_hot_cowling.eps}
\caption{{\em Perturbed, magnetized hot NS with the SHT EoS in a fixed spacetime}.
{\bf Top:} The (normalized) central density of the star
and central magnetic field strength as a function of time.
{\bf Bottom:} The Fourier power spectral density as a function
of frequency.
Leakage results for this star are shown Fig.~\ref{fig:luminosities_gshen_hot_cowling}.}
\label{fig:rhoBLc_gshen_hot_cowling}
\end{figure}

\begin{figure}[h]
\centering
\includegraphics[height=7.0cm,width=8.5cm,angle=0]{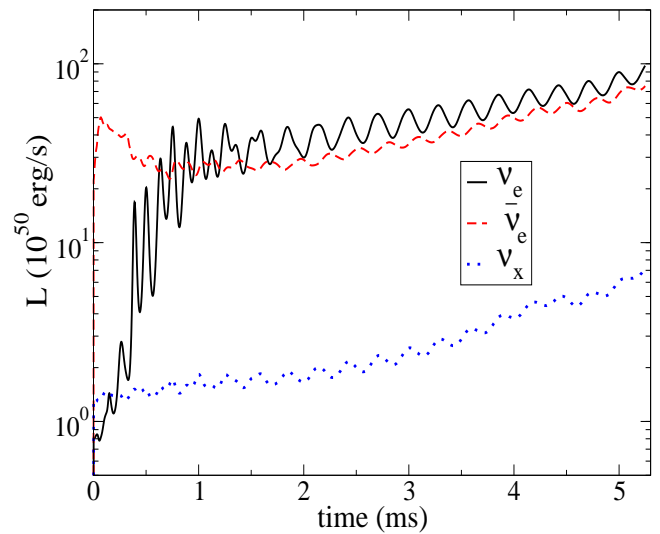} \\
\caption{{\em Perturbed, magnetized hot NS with the SHT EoS in a fixed spacetime}.
The luminosities of the different neutrino species are displayed
as a function of time. Notice that, after the initial
transient of $\approx 1$ ms, all luminosities rise steadily.
We attribute this to the increasing temperature near the star's surface
due mainly to numerical diffusion and to the evolution of the atmosphere.}
\label{fig:luminosities_gshen_hot_cowling}
\end{figure}

Here we study a hot star, chosen to match one 
already studied in Ref.~\cite{Galeazzi:2013mia}, also with
a central density $\rho_c = 9.3\times10^{14}\,{\rm g/cm^3}$.
In particular, we choose a star with constant entropy 
of $s=1\,k_B/\mathrm{baryon}$ in $\beta$-equilibrium using the SHT EoS,
leading to a temperature of $T \approx 30$\,MeV at the center, which decreases
towards the surface. 
We let discretization error serve as the only perturbation. 

We evolve in the Cowling approximation
by freezing the spacetime in order to compare
the normal frequencies with the ones obtained from the linearized
system. The star is given an initial, poloidal, magnetic field with
maximum strength $5 \times 10^{14}\,{\rm G}$, and we allow the star to cool via
neutrino emission as described by the leakage scheme.
The central values of the density and the magnetic field
are plotted in Fig.~\ref{fig:rhoBLc_gshen_hot_cowling}.
The frequencies, summarized in Table~\ref{tab:stars}, are in good
agreement with those calculated in Ref.~\cite{Galeazzi:2013mia}.

We also display the neutrino
luminosities for each species in
Fig.~\ref{fig:luminosities_gshen_hot_cowling}. 
During the first millisecond, 
both the $\nu_e$ and $\bar{\nu}_e$ luminosities vary significantly
as the star achieves its new equilibrium configuration for the chosen
numerical grid. In particular, we find a
mixing of the neutron-rich matter and an increase of the temperature
near the stellar surface due to
(i)~the fluid evolution
outside the star (i.e., with thermal ejections of stellar material,
shock heating and accretion of the atmosphere), (ii)~the internal normal
oscillations of the star, and (iii)~numerical diffusion. These effects
are enough to drive the $Y_e$ away from its original $\beta-$equilibrium
configuration and to produce a slow and steady rise of all the neutrino
luminosities after $t \approx 2$\,ms.
Before the steady rise, the $\nu_e$,
$\bar{\nu}_e$, and $\nu_x$ luminosities are $\approx 3\times
10^{51}$\,erg/s, $\approx 2\times10^{51}$\,erg/s, and $\approx 
0.2\times10^{51}$\,erg/s, respectively.
Here, we include all four
heavy-lepton species in the $\nu_x$ luminosity.  

Our results show a much lower neutrino luminosity than those of
\cite{Galeazzi:2013mia} for the same model, who at the stationary state
have luminosities of $\approx 30\times 10^{51}$\,erg/s, $\approx 70\times
10^{51}$\,erg/s, and $\approx  15\times 10^{51}$\,erg/s, for $\nu_e$,
$\bar{\nu}_e$ and $\nu_x$, respectively.
However, high resolution (50\,m) tests with
this same stellar configuration in GR1D both with this identical
leakage scheme \cite{O'Connor:2009vw} and with a two-moment closure neutrino transport scheme \cite{O'Connor:2012am}
suggest that our luminosities are reasonable and indicate that these large
differences may arise due to the treatment of low density regions,
which seems to be crucial in this particular problem. With GR1D's leakage scheme we
achieve luminosities at the stationary state of $\approx 0.8\times 10^{51}$\,erg/s,
$\approx  0.4\times10^{51}$\,erg/s, and $\approx 0.7\times 10^{51}$\,erg/s.  With
GR1D's two moment neutrino transport, we determine hydrostatic
neutrino luminosities of $\approx 0.3\times10^{51}$\,erg/s,
$\approx 0.5\times10^{51}$\,erg/s, and $\approx 0.8\times10^{51}$\,erg/s for
$\nu_e$, $\bar{\nu}_e$, and $\nu_x$, respectively. We see lower
luminosities in spherical symmetry because of the reduced numerical
diffusion and the inability to capture convection in the atmosphere
region.

We are able to attribute all the observed differences in the neutrino
luminosities in our 3D simulations and the GR1D simulations to the
outer $\approx 1-2$\,km of the star.  Interior to this, the leakage
quantities are initially essentially identical and vary only
slightly. In the interior region, the grey radial luminosity determined with GR1D's
leakage scheme is in very good agreement with the energy summed
neutrino luminosity from GR1D's neutrino transport. The outer regions
are much harder to accurately capture with a leakage scheme as both
the neutrinospheres and the region where the dominant neutrino leakage
contributions transitions from diffusion to free emission, occur very
close to the steep density gradient of the neutron star's surface. For
this reason, we express caution when interpreting these luminosities.

\begin{figure}[h]
\centering
\includegraphics[height=9.0cm,width=9.5cm,angle=0]{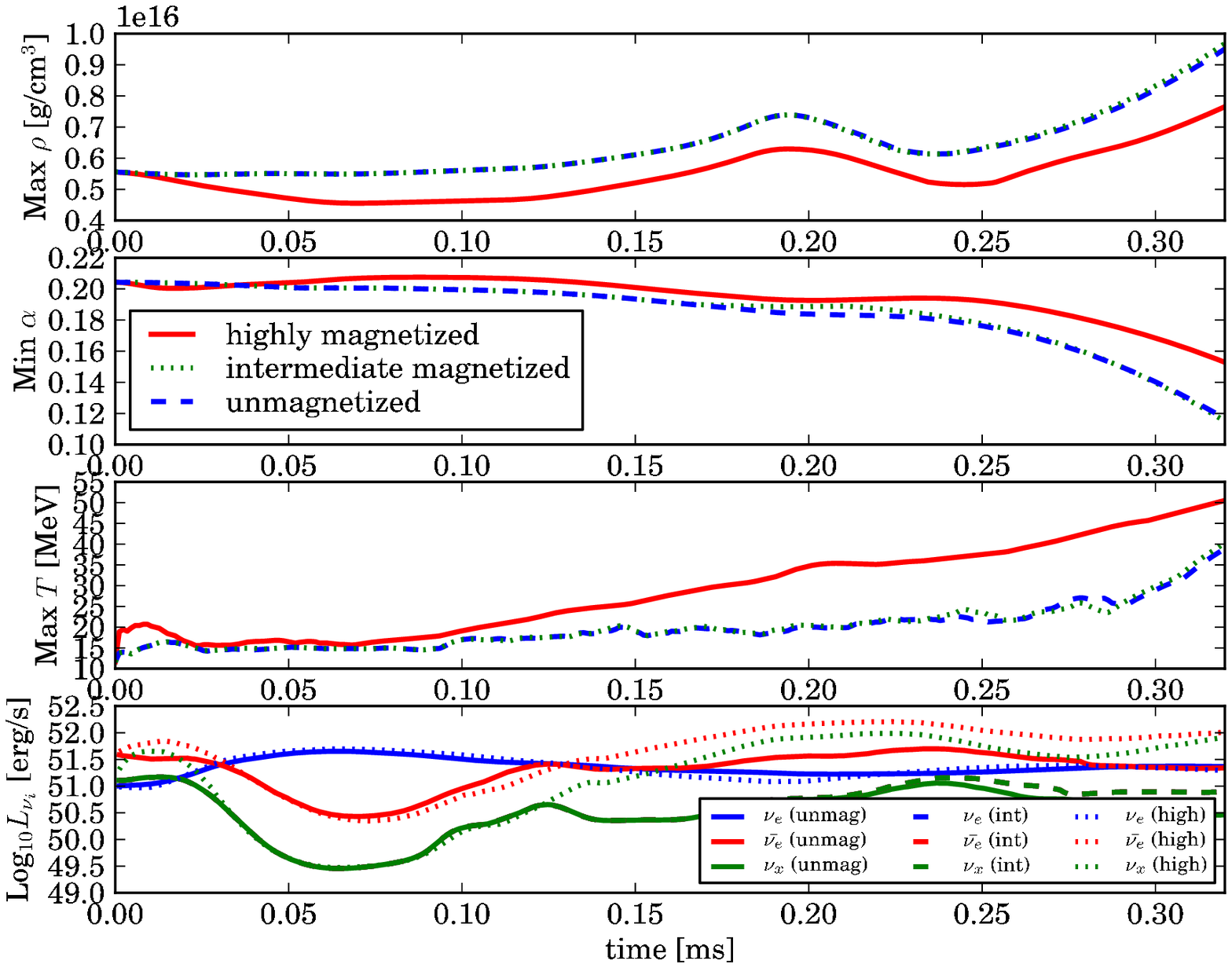} 
\caption{{\em Hot, dense, rapidly rotating star.}
The maximum density, minimum lapse, maximum temperature and total neutrino luminosities are shown versus time for the collapse of an unstable, hot, dense, rapidly rotating star. Three evolutions are contrasted, one with no magnetization (blue dashed) one with a very large initial field (red solid), and one with an intermediate magnetization (green dotted).
The intermediate star is nearly identical to the unmagnetized star except for its
late-stage neutrino luminosity.
The magnetized star has an initial  magnetic field with a maximum of $4.1 \times 10^{16}$ G whereas
as the most magnetized case begins with a field one hundred times larger.
Pictures of the star in its late stage are shown in Fig.~\ref{fig:hotdense2D}.
}
\label{fig:hotdense1D}
\end{figure}
\begin{figure}[h]
\centering
\includegraphics[width=7.6cm,angle=0]{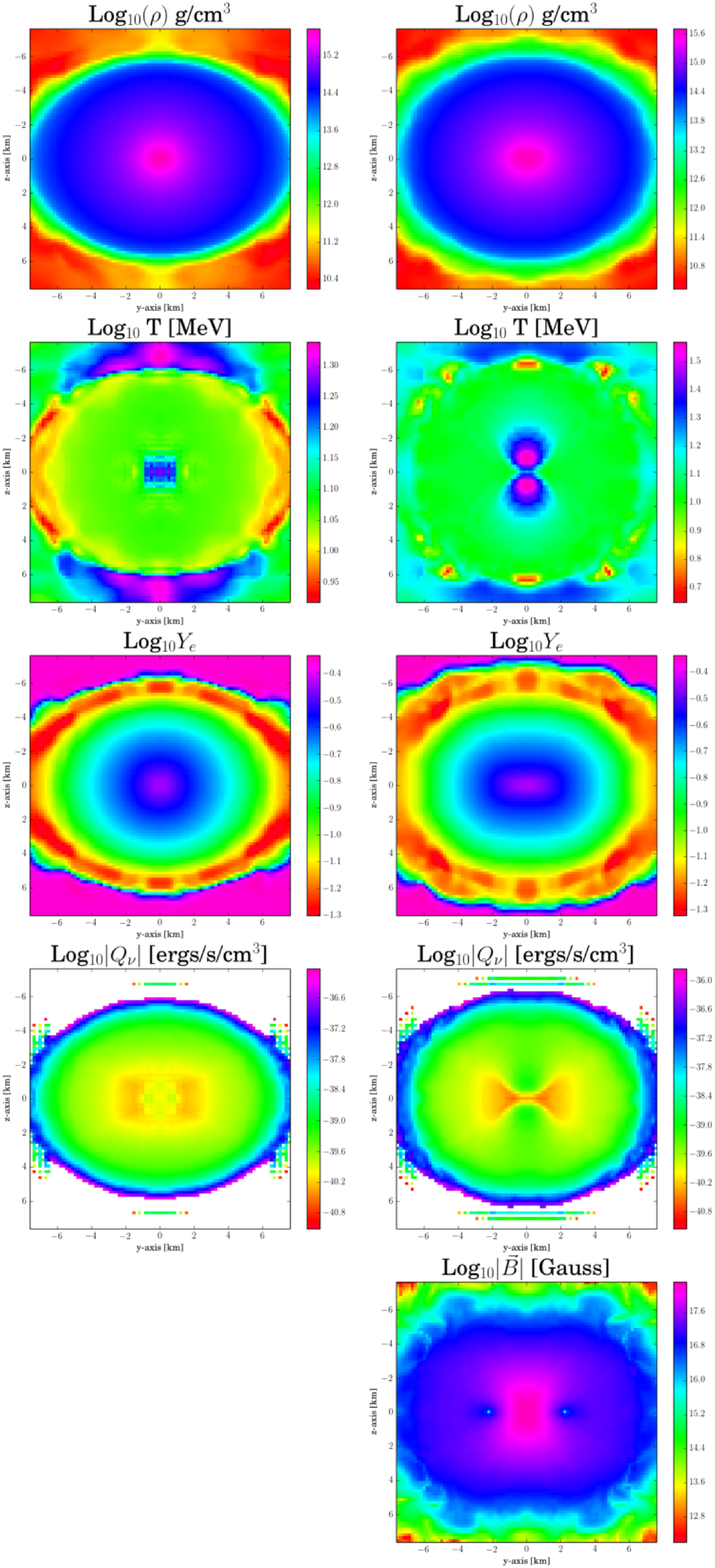}
\caption{{\em Hot, dense, rapidly rotating star.}
Snapshots of various quantities on a meridional plane are shown at
$t=0.25$ms for the same unmagnetized star (left) and highly magnetized star (right) as shown in Fig.~\ref{fig:hotdense1D}. From top to bottom are shown the density, temperature, electron fraction, total neutrino emissivity, and the magnitude of the magnetic field.
The square central region in the temperature (and concomitantly the emissivity) 
of the unmagnetized star appears related to the squared boundaries.
}
\label{fig:hotdense2D}
\end{figure}

In an effort to better understand the effects of both magnetic field and
neutrino cooling, we study the evolution of a
hot ($12$ MeV),
dense ($5.6\times 10^{15}$ g/cm${}^3$),
rapidly spinning ($1500$ Hz) neutron star using the HS EoS.
For such a large central density and mass (baryonic mass $2.1 M_\odot$), the star is unstable to collapse providing
a dynamic solution to study. 
We consider the star with no magnetic field
and contrast it when the star is given a very strong magnetic field.
We consider two magnetized cases parametrized by a maximum initial magnetic field strength of
$4.1 \times 10^{16+n}$ with $n=0,2$. The low value ($n=0$) could arise dynamically in the hypermassive 
neutron star resulting from binary merger~\cite{Price:2006fi,Anderson:2008zp,2010A&A...515A..30O,Zrake:2013mra}.
The high value ($n=2$), while unrealistically high, allows us to explore
strong magnetic fields which can have a significant effect on the pressure and structure of the star and, consequently,
on the neutrino production. 
Notice that we neglect these effects in the construction of the initial data, adding the magnetic field to the star as a 
``seed.''  Nevertheless, these evolutions serve primarily to
explore some of the possible differences and the robustness of the code, in addition to assessing at which
level such magnetizations can affect the dynamics.

Fig.~\ref{fig:hotdense1D} shows various quantities as functions of time for both
unmagnetized and magnetized evolutions. In Fig.~\ref{fig:hotdense2D}, we show
plots of certain fields at a late time during the collapse.
As shown in the figure, only
unreasonably high magnetizations
significantly  changes the neutrino production. 
Magnetic fields do not couple strongly to neutrino cooling in realistic scenarios and instead
affect neutrino results through effects to stellar structure and temperature.
Notice however, that in binary mergers the magnetic field can significantly affect the distribution of 
material and the transport of angular momentum of the merger remnant. And these effects can, in turn,
impact neutrino production and luminosity.

\subsection{Binary Neutron Stars}

\begin{figure}[h]
\centering
\includegraphics[width=8.5cm,angle=0]{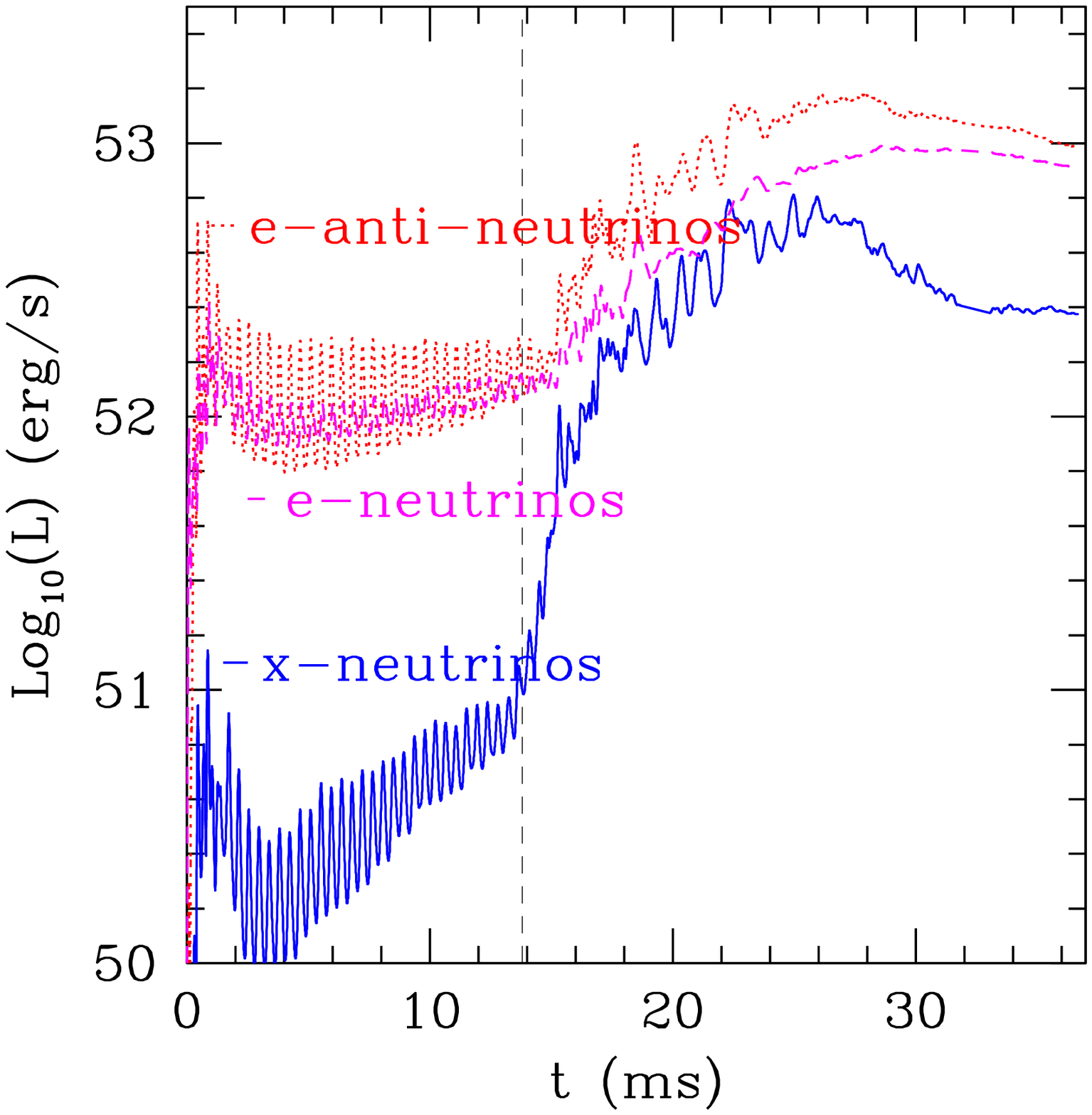}
\caption{{\em Neutrino luminosities for the merger of a binary neutron star system.} The dashed line at $t=13.8$ms denotes when the stars first touch.
Note the rapid growth of the luminosities for all species after this time.
Snapshots of the binary at a few of these late times are shown in Fig.~\ref{fig:bns}.
} 
\label{fig:bnslums}
\end{figure}
\begin{figure*}
\centering
\includegraphics[width=0.475\textwidth,angle=0]{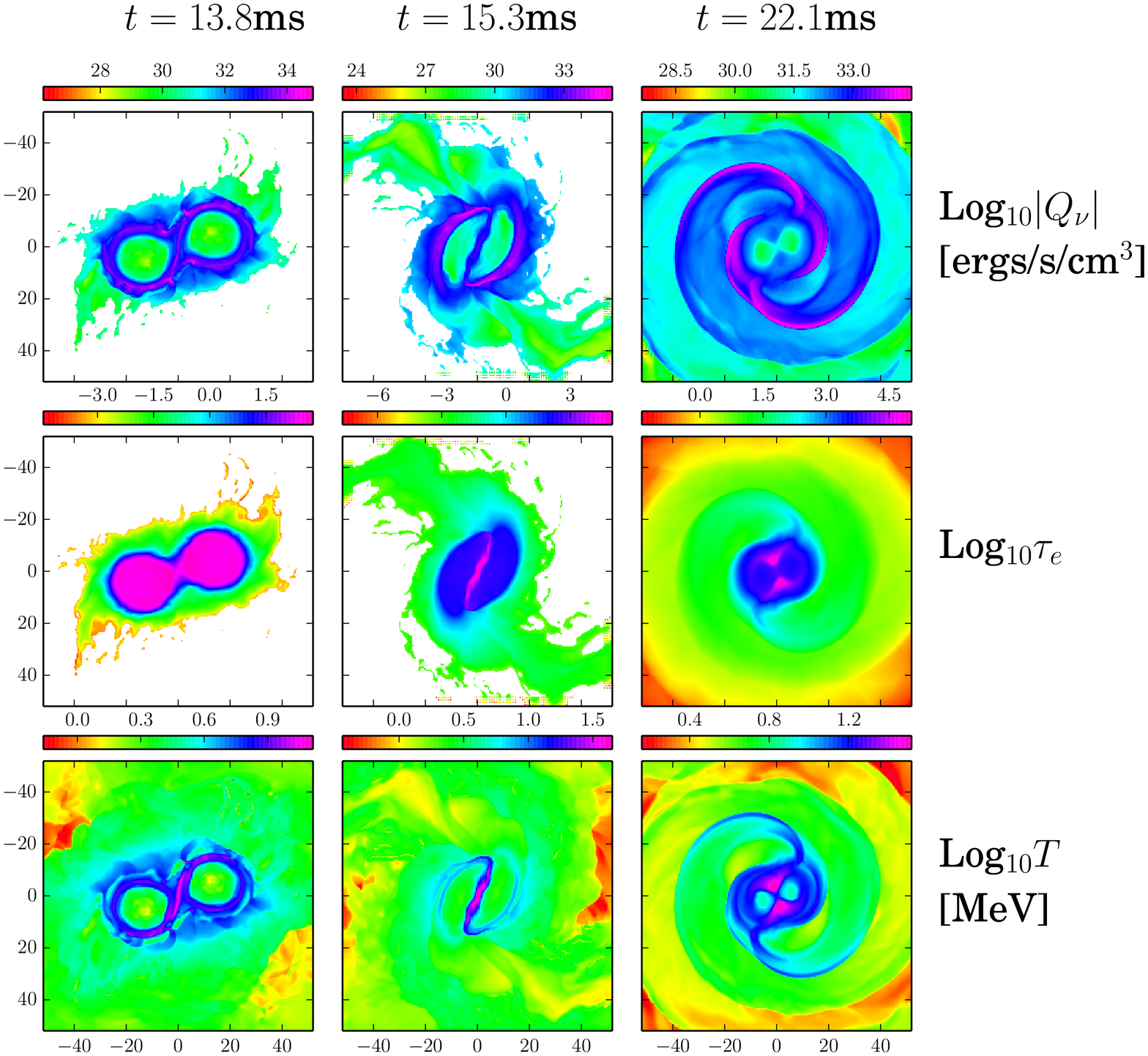}
\includegraphics[width=0.475\textwidth,angle=0]{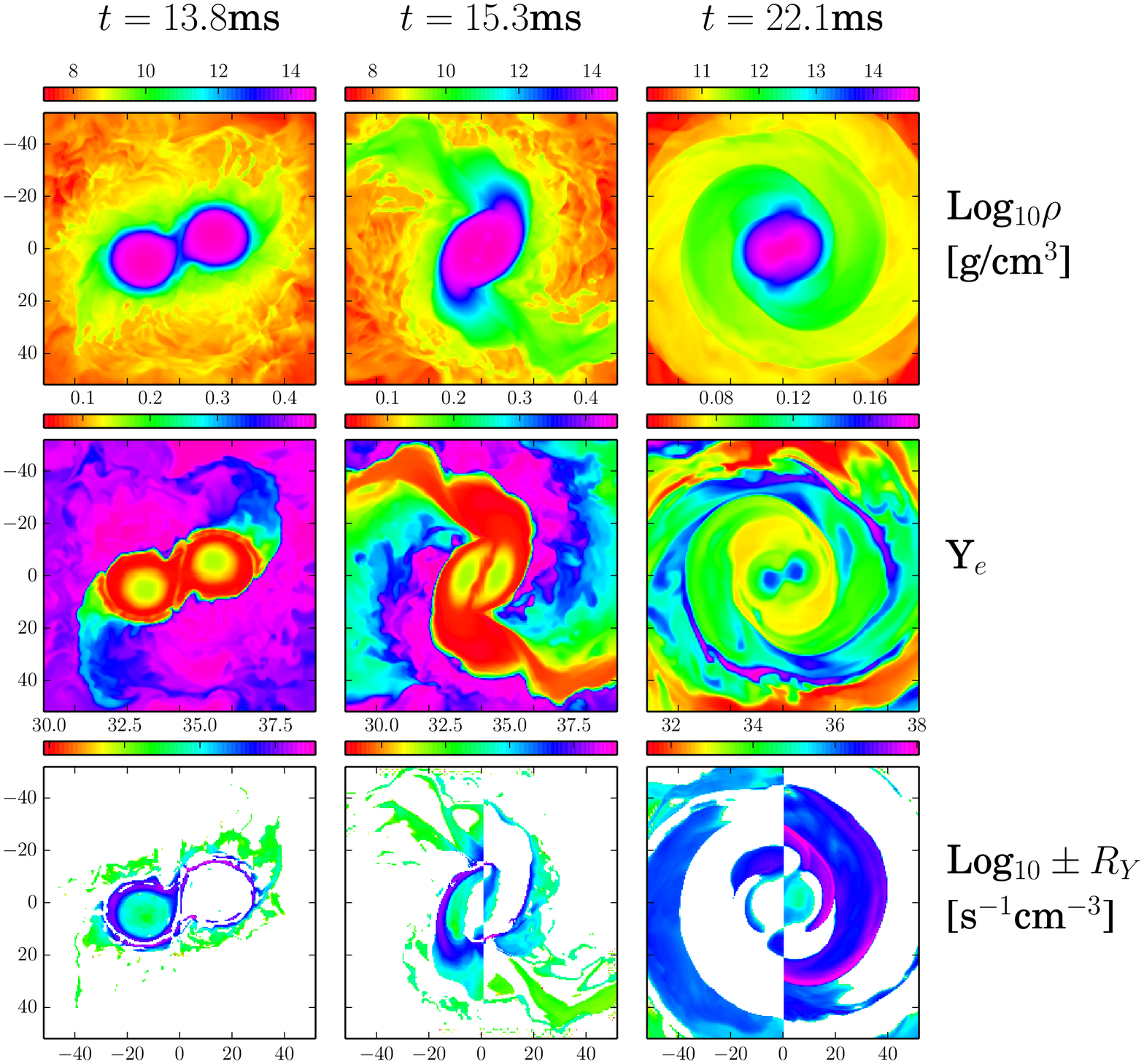}
\caption{{\em Snapshots of the binary merger looking down upon the orbital plane.}
The total neutrino emissivity, electron-neutrino optical depth, and temperature are shown at
three times ($t=13.8$ms, $15.3$ms, and $22.1$ms) increasing from left to right.
For the plots of $R_Y$, the right side ($x>0$) of the plot shows positive values (net electron antineutrino emission) whereas the
left side ($x<0$) shows negative values (net electron neutrino emission).
The first time shows when the stars first touch and
the middle time shows the system when the temperature reaches its
maximum. The final column shows the remnant at $t=22.1$ms, a while after merger when it has settled down to a rotating,
hot ``dumb-bell'' with spiral arms of emissive material.
The luminosities for each neutrino species as functions of time are shown in Fig.~\ref{fig:bnslums}.
} 
\label{fig:bns}
\end{figure*}

The inspiral and merger of two neutron stars is a significant test of
a code such as this. The inherent asymmetry of the problem tends
to ``excite'' many, if not all, terms in the equations. Resolving
both the motion
of two compact objects as well as the large
gradients at their  surfaces requires significant resources.
The merger itself is a very dynamic process
with a large range of densities.

We consider here just the late stage (roughly the last $4.5$ orbits)
of a single binary evolved with just a single EoS.  We have evolved this binary
with three different resolutions and find convergent results. The preliminary results presented 
here arise from our highest resolution which has a finest grid spacing in all directions of $460$m. 
This resolution is enough to capture the main aspects of the dynamics, though we note higher resolutions
are needed for a more detailed study of the system.

We include neutrino cooling during the inspiral as a test of our
leakage scheme even though for times much before merger one expects very little emission.
Above we show that the computation of the optical depth
tracks the stars appropriately in Fig.~\ref{fig:optdepthBNS}.
Here we show results for a binary constructed using
{\sc Lorene} with the HS EoS for which each star has baryonic mass $M_B=1.49\,M_\odot$ and
equatorial radius $R \approx 14.5$\,km  and temperature $0.01$\,MeV. The binary has an initial separation
$a=45$\,km, a total ADM mass $M_{\rm ADM}= 2.74\,M_\odot$, and
an orbital angular velocity $\Omega=1796$\,rad\,s$^{-1}$. The electron fraction is set so that
the stars are initially in $\beta$-equilibrium. Neutron star binaries are 
expected to be rather cool, but higher temperatures will be reached during merger.
We have run this binary at three resolutions and find that the hydrodynamical and gravitational dynamics  are
convergent. 

%
%

In Fig.~\ref{fig:bnslums} we show the integrated luminosities for each
neutrino species as a function of time.
After an initial transient, the luminosities for both electron-neutrinos
and electron-anti-neutrinos become roughly comparable while the heavy-lepton
neutrino types are much less luminous. The large oscillations in the
luminosities are an unfortunate artifact of an inconsistency between
the temperature provided to the initial data solver and that provided
to the evolution code.
This inconsistency acts as a very large perturbation
to the binary resulting in oscillations in the initial density that naturally
induce oscillations in the other fields, but otherwise does not affect the dynamics
of the orbiting stars.
A simulation such as this assumes no symmetries and therefore
places no restrictions on the dynamical physical modes that the merger can
excite.
Because an evolution of the full, three-dimensional domain requires close to one 
hundred-thousand cpu-hours, we defer for our follow-up work simulations with better 
initial data for which preliminary results show much smaller oscillations.

Once the stars touch, the temperature and concomitant neutrino luminosity increase
for all species. Interestingly, the most dominant radiative neutrino flavor
is the electron antineutrino. 
This dominance has already been 
observed in other binary neutron star 
mergers~\cite{Sekiguchi:2011zd,Sekiguchi:2012uc}
and also neutron star--black hole mergers~\cite{Deaton:2013sla}, and 
is due to neutron rich material being shock heated and
decompressed~\cite{1997A&A...319..122R}.  This
neutron rich, low density, hot material is initially far below the
new $\beta-$equilibrium value of $Y_e$ and will preferentially emit
electron antineutrinos until it is reestablished.  
This can be seen by inspection of the bottom row on the right side of Fig.~\ref{fig:bns},
which shows the lepton number matter source term. On the left
  half ($x<0$) of each panel we show (in color-log scale) regions of
  net positive lepton number emission and mask out regions of net
  negative lepton number emission. While on the right half ($x>0$)
  we only show regions of net negative lepton number
  emission and mask out positive lepton number emission.  In the
  rightmost column, at 22.1\,ms, the largest contribution to the
  antineutrino emission (again, $x>0$) occurs at the leading edges of
  the tidal tails where $T\approx 15$\,MeV, $\rho \approx 
  10^{12}$\,g\,cm$^{-3}$ and $Y_e\approx 0.1$.  
This can be understood by noting that for this temperature, density,
and EoS, we have that $Y_{e,\beta-\mathrm{equil}}\approx0.29$,
significantly above the instantaneous value of $Y_e \approx 0.1$. In
contrast, the dominant region of electron neutrino number emission,
albeit small, is outside of the dominant tidal tail region (for radii
$\gtrsim 40$\,km). This matter is colder and less dense, characterized
by $T\approx 2.5$\,MeV, $\rho\approx (1-2) \times
10^{11}$\,g\,cm$^{-3}$ and $Y_e \approx 0.1-0.15$.  Since the optical
depth is low for this region, the equilibrium $Y_e$ that the system
will tend to is the $Y_e$ where neutrino number emission balances
antineutrino number emission rather than the $\beta-$equilibrium value,
which assumes there exists a population of trapped neutrinos and
antineutrinos that can undergo capture on neutrons and protons. For
these conditions, $Y_{e,\mathrm{rate\ equil}} \approx 0.05$,
hence we generally expect dominant electron neutrino emission.

Our results are largely consistent with past work for similar mass
binaries, indicating that despite differences in the adopted
approaches, simulations predict a rather robust behavior of neutrino
luminosities $\approx 10^{53}$\,ergs\,s$^{-1}$. The dominant
contribution is provided by electron antineutrinos, followed with a
somewhat lower luminosity by electron neutrinos. 
The heavy-lepton
  neutrino luminosity is roughly half those of the light
  species  for the roughly $15$\,ms immediately after the merger,
  but decays considerably
  afterwards as the disk settles and cools.
The faster decline of the
  heavy-lepton luminosities relative to the electron type species is
  due to the different emission processes. Heavy-lepton neutrino
  emission is dominated by electron-positron annihilation, which has a
  stronger temperature dependence than charged-current processes which
  dominate the electron-type neutrino emission.

Snapshots corresponding to the merging binary are shown in Fig.~\ref{fig:bns}
illustrating, at representative times after merger, the behavior of: energy sink
($Q_{\nu}$), electron 
neutrino  optical depth ($\tau_e$), temperature ($T$), density ($\rho$), 
electron fraction ($Y_e$) and lepton source ($R_Y$). This figure illustrates
several important features. First, the energy sink/lepton sources and temperature 
are tied to surface, shearing regions and tidal tails. Second, the optical
depth tracks, as expected, the density behavior. Third, the electron fraction
shows regions above/below beta-equilibrium which induce stronger production
of electron neutrinos/antineutrinos. Fourth, the merger gives rise to a significant
increase in the temperature,  reaching a peak value of roughly $45$~MeV. 
Notice that for this low total-mass binary, the HMNS
does not show indications of impending collapse for the time of
this simulation (roughly 25ms after merger) implying a longer lived remnant.
We defer to future work a systematic analysis of the dependence on total mass,
EoS and magnetic field effects on the onset of black hole formation, 
as well as other properties of the merger remnant, mass outflow, and 
resulting nucleosynthesis.

\section{Conclusions}
\label{sec:conclusions}
We have described our implementation of a general relativistic, magnetohydrodynamics code
that uses realistic, tabulated  equations of state as well as
accounts for neutrino cooling through a leakage scheme. In particular, we
have presented a new method to obtain the optical depth along with a test illustrating
that the optical depth is accurately integrating the opacity\footnote{At the completion of this work,
an independent, related approach has
been presented in~\cite{Perego:2014qda}.}. With this implementation
we have studied the effects magnetic fields can have on the collapse of a rapidly spinning,
hot, dense star with neutrino cooling. We also presented first applications of this
code to a binary neutron star system obtaining, in particular, the neutrino luminosity
induced by the merger. For both single star cases and binaries, the calculated values are 
in agreement with recent work~\cite{Sekiguchi:2012uc,Galeazzi:2013mia} as well as with
results obtained with a core collapse code~\cite{O'Connor:2009vw}.

%
%
\vspace{0.5cm}
\begin{acknowledgments}
It is a pleasure to thank C. Ott, T. Janka, S. Rosswog,
L. Caballero, F. Galeazzi, and A. Burrows for discussions about neutrino
effects and techniques, as well
as our long time collaborators E. Hirschmann, P. Motl and M. Ponce.
This work was supported by the NSF under grants PHY-0969827 \& PHY-1308621~(LIU),
PHY-0969811 \& PHY-1308727~(BYU), NASA's ATP program through grant NNX13AH01G,
NSERC through a Discovery Grant (to LL) and CIFAR (to LL).
C.P acknowledges support by the Jeffrey L.~Bishop Fellowship.
Research at Perimeter Institute is supported through Industry Canada and by the Province of Ontario
through the Ministry of Research \& Innovation.  Computations were
performed at XSEDE and Scinet.
\end{acknowledgments}

\bibliographystyle{utphys}
\bibliography{paper}

\end{document}